\documentclass{sig-alternate}

\usepackage{PageSetting}
\usepackage{wrapfig,epsfig}

\usepackage{algorithm}
\usepackage{algorithmic}

\newtheorem{fact}{Fact}

\Alph{algorithm}

\usepackage{times}

\def\done{\hspace*{\fill} $\framebox[2mm]{}$}

\title{Small Count Privacy and Large Count Utility in Data Publishing}

\author{Ada Wai-Chee Fu$^1$, Jia Wang$^1$, Ke Wang$^2$, Raymond Chi-Wing Wong$^3$
\\
\\
\normalsize
$^1$Department of Computer Science and Engineering, Chinese University of Hong Kong\\
\normalsize
$^2$Department of Computer Science, Simon Fraser University \\
\normalsize
$^3$ Department of Computer Science and Engineering, the Hong Kong University of Science and Technology \\ 
\large
adafu,jwang@cse.cuhk.edu.hk,  wangk@cs.sfu.ca, raywong@cse.ust.hk 
\\
}

\begin{document}

\maketitle

\begin{sloppy}
\begin{abstract}
While the introduction of differential privacy has been a major breakthrough in
the study of privacy preserving data publication, some recent work has
pointed out a number of
cases where it is not possible to limit inference about individuals.
The dilemma that is intrinsic in the problem is the simultaneous requirement of
data utility in the published data.
Differential privacy does not aim to protect
information about an individual that can be uncovered even without the participation
of the individual. However, this lack of coverage may violate
the principle of individual privacy.
Here we propose a solution
by providing protection to sensitive information, by which we refer to
the answers for aggregate queries with small counts.
Previous works based on $\ell$-diversity can be seen
as providing a special form of this kind of protection.
Our method is developed with another goal which is
to provide differential privacy guarantee,
and for that we introduce a more refined
form of differential privacy to deal with certain practical issues.
Our empirical studies show that our method can preserve better utilities
than a number of state-of-the-art methods although these methods do
not provide the protections that we provide.

\end{abstract}

\section{Introduction}

The ultimate source of the problem with
privacy preserving data publishing is that we
must also consider the utility of the published data.
The problem is intriguing to begin with because we have a pair of seemingly
contradictory goals of utility and privacy.
Whenever we are able to provide some useful information with the published data,
there is the question of privacy breach because of that information.


The statistician Tore Dalenius advocated the following
privacy goal in \cite{Dal77}: {\it Anything that can be learned about
a respondent from the statistical database should be learnable without
access to the database.}
To aim for this goal,
some previous works have considered the
approach where prior and post beliefs about an individual are to be
similar \cite{EGS03,RSH07,agrawal05}.
As discussed in \cite{D08}, this privacy goal may be self-contradictory
and impossible
in the case of privacy preserving data publication.
The goal of the published data is for a receiver to know something
about the population, it is by definition that the receiver can discover
something about an individual in the population, and the receiver could happen
to be the adversary.
Due to the seeming impossibility of the above goal, research in differential privacy
moves away from protecting the information about a row in the data table that
can be learned from other rows \cite{BDMN05}.
The argument is that such information is derivable without the participation of the corresponding individual in the dataset, and hence is not under the
control of the individual. However, although not under
the individual's control, such information could nevertheless
be sensitive. 

An important goal of our work here is to show that it is
possible to protect sensitive information that
can be acquired from the published dataset
provided that the data publisher,
with control over the dataset, can
act on behalf of each individual. The principle
of protecting individual privacy may dictate that
the publisher either provides this kind of protection or does not publish the data. It is desirable that on top of ensuring that the participation of a user
makes little difference to the results of data analysis, the publisher also guarantees protection for sensitive information that can be derived from the published dataset,
with or without the data of the individual involved. Such a solution is our goal.
While Dalenius's original goal may be impossible, it
is also an overkill.
A ``relaxed'' goal suffices:
{\it Anything ``sensitive'' that can be learned about
a respondent from the statistical database should be learnable without
access to the database.} The obvious question is what should be considered
sensitive. We provide a plausible answer here.

Let us consider an example given in \cite{D11},
where a dataset $D'$ tells us that almost everyone involved in a dataset
has one left foot and one right foot. We would agree that
knowing with high certainty that a respondent is two footed
from $D'$ is not considered a problem since almost everyone is two footed.
Note that even if an individual does not participate in the data collection,
the deduction can still be made based on a simple assumption of an i.i.d.
data generation.
Differential privacy and all of the proposed privacy models
so far do not exclude the possibilities of
deriving information of such form. In fact, by definition of data
utility, such a derivation should be supported.
This example is not alarming since it involves a large population.
However there will be cases where the
 information becomes sensitive and requires protection.
Let us consider a medical data set. Suppose lung cancer is not a
common disease. Also suppose
there are only five females aged 70, with postal code 2980
and all of them have lung cancer, the linkage of the corresponding
(gender, age, postal code) with lung cancer in this case is 100\%; if we maintain high utility
for accurately extracting such information or concepts,
the privacy of the five females will be compromised.
The reason why this is alarming is because accurate answers to queries of small counts can disclose highly sensitive information.
A problem with many existing techniques lie in
non-discriminative utilities for all concepts.
We propose to consider discriminative utilities
which are based on
the population sizes: queries involving large populations
can be answered relatively accurately while queries with a very small
population base should not.
A similar idea is found in the literature of
security for statistical databases
\cite{AdamSecurityControl89,TYW84,Fellegi72,HM76,Denning80}
(see Section \ref{related} on related work).

\label{issues}

Protecting queries of small counts is implicit
in many previous works.
For example, the principle of $\ell$-diversity \cite{l-diversity}
essentially protects against accurate answers to queries about
the sensitive values of individuals, which may become
small count queries given that the adversary has knowledge
about the non-sensitive values of an individual and
therefore is capable of linkage
attack \cite{Sweeney97,samarati-protecting}.
We shall show that our mechanism provides better protection
when compared to $\ell$-diversity approaches.


\label{selectQ}

Our major contributions are summarized as follows.
We point out the dilemma that
utility is a source of privacy breach, so that on top of differential
privacy we must also protect sensitive information that can be derived from
the published data.
We propose a mechanism for
privacy preserving data publication which provides three lines
of protection: (1) differential privacy to protect information
that may be attained from the data
 of an individual tuple, (2) protection
for concepts with small counts which can be derived from
the entire published data set, and
(3) a guarantee that the published data does not narrow down the
set of 
possible sensitive values for each individual. We enforce a stronger
$\epsilon$-differential privacy guarantee by setting $\epsilon = 0$.
We support discriminative
utilities so that concepts with large counts
can be preserved.
While $\ell$-diversity methods are vulnerable to 
adversary knowledge that eliminates $\ell-1$
possible values, our method is resilient to such attacks.
We have conducted experiments on
a real dataset to show that our method provides better utilities
for the large sum queries than several state of the art methods
which do not have the above guarantee.

The rest of the paper is organized as follows.
In Section 2, we revisit $\epsilon$-differential privacy
for non-interactive database sanitization. We point out issues about
$\epsilon$ and about known presence. Then we introduce our model of
$\ell'$-diverted zero-differential privacy.
Section 3 describes a first attempt of a solution using an existing
randomization method, we show that this method cannot guarantee zero-differential
privacy. Section 4 describes our proposed mechanism $A'$
which generates $D'$.
Section 5 is about count estimation
given $D'$. Section 6 shows that
mechanism $A'$ supports high utility for large counts and high inaccuracies for
small counts. Section 7 is about multiple attribute aggregations.
Section 8 is a discussion about auxiliary knowledge that may be possessed by the adversary.
Section 9 reports on the empirical study. Related works are
summarized in Section 10 and we conclude in Section 11.

\section{$\ell$'-diverted Privacy}

Our proposed method guarantees a desired form of
differential privacy with the additional
protection against the disclosure of sensitive information of small counts.
In this section we shall introduce our definition of privacy guarantee
based on differential privacy.
First we examine some relevant definitions from previous works.
%
%
%
%
%
%
The following
is taken from \cite{BLR08}.

\begin{definition}[${\cal A}(D)$ and $\epsilon$-differential privacy]
For a database D, let $\cal A$ be a database sanitization mechanism,
we will say that ${\cal A}(D)$
induces a distribution over outputs.
We say that mechanism ${\cal A}$
satisfies $\epsilon$-differential privacy if for all neighboring databases
$D_1$ and $D_2$ (i.e. $D_1$ and $D_2$ differ in at
most one tuple), and for all sanitized outputs $\hat{D}$,
$Pr[ {\cal A}(D_1) = \hat{D} ] \leq e^{\epsilon} Pr[ {\cal A}(D_2) = \hat{D} ] $
\done
\label{defn0}
\end{definition}

The above definition
says that for any two neighboring databases, the probabilities that ${\cal A}$
generates any particular dataset for publication are very similar.
However, there are some practical problems with this definition.

\subsection{The problem with $\epsilon$}

In $\epsilon$-differential privacy, the parameter $\epsilon$ is public.
The sanitized data is released to the public, and the public
refers to a wide spectrum of users and applications. It is not at all
clear how we may have the parameter $\epsilon$ decided once and for all.
In \cite{D11}, it is suggested that we tend to think of $\epsilon$ as,
say, 0.01, 0.1, or in some cases, $\ln 2$ or $\ln 3$.
Evidently the value can vary a lot. For example, for the above suggested values,
$e^\epsilon$ ranges from 1.01, 1.105 to 2 and 3.

A second problem with the setting of $\epsilon$
is that it may compromise privacy.
Suppose that for all pairs of neighboring datasets $D_1$ and $D_2$, where
$D_2$ contains $t$ while $D_1$ does not,
$Pr[{\cal A}(D_1)=\hat{D}]$ is 1/3, while 
$Pr[{\cal A}(D_2)=\hat{D}]$ is 1.
If we set $\epsilon$ to $\ln 3$, then
 $\epsilon$-differential privacy is satisfied,
but the existence of $t$ can be estimated with 75\%
confidence.

The above concerns call for the elimination of $\epsilon$.
We can do so by setting $\epsilon$ to zero.
This is in fact the best guarantee since it means that there is no
difference between $D_1$ and $D_2$ in terms of the probability of
generating $D'$. We shall refer to this guarantee as
{\it zero-differential privacy}.

\subsection{The issue of known presence}

While the initial definition of differential privacy aims at hiding the
presence or absence of an individual's data record, it is often the case
that the presence is already known. As discussed in \cite{D11},
in such cases, rather than hiding the presence, we wish to hide certain
values in an individual's row. We shall refer to such values that need to
be hidden as the sensitive values.
The definition of differential privacy need to
be adjusted accordingly. The phrase "differ in at most one tuple"
in Definition \ref{defn0}
can be converted to "have a symmetric difference of at most 2".
This is so that in two datasets $D_1$ and $D_2$, if only the data for one individual
is different, then we shall find two tuples in the symmetric difference of
$D_1$ and $D_2$. The two tuples are tuples of the same individual in the two
datasets, but the sensitive values differ.
However, with this definition,
the counts for sensitive values in $D_1$ or $D_2$ would
deviate from the original data set $D$. For a neighboring database, we
prefer to preserve as much as possible the characteristics in $D$.
In the following subsection, we introduce a definition of differential
privacy that addresses the above problems.

\subsection{$\ell$'-diverted zero-differential privacy}

\label{nondiff}

Given a dataset (table) $D$ which is a set of $N$ tuples,
the problem is how to generate sensitive values for the tuples in
$D$ to be published in the output dataset $D'$.
We assume that there are two kinds of attributes in the dataset,
the non-sensitive attributes ($NSA$) and a sensitive attribute
($SA$) $S$. Let the domain of $S$ be $domain(S) =$ $\{s_1, ..., s_m\}$.
We do not perturb the non-sensitive values but we
may alter the sensitive values in the tuples to ensure privacy.
We first introduce our definition of neighboring databases which preserves
the counts of sensitive values, and we minimize the moves by swapping
the sensitive values of exactly one arbitrary pair of tuples with different
sensitive values. In the following we use $t.s$ to denote the value of the 
sensitive
attribute of tuple $t$.


\begin{definition}[neighbor w.r.t. $t$]
Suppose we have two databases $D_1$ and $D_2$
containing tuples for the same set of individuals,
and $D_1$ and $D_2$ differ only at two pairs
of tuples, $t$, $\breve{t}$ in $D_1$ and $t', \breve{t'}$
in $D_2$.
Tuples $t$ and $t'$ are for the same individual,
and $\breve{t}$ and $\breve{t'}$ are for another individual,
with $t.s \neq \breve{t}.s$,
$t.s = \breve{t'}.s$, and $\breve{t}.s = t'.s$.
Then we say that $D_2$ is a
neighboring database to $D_1$ with respect to $t$.
\end{definition}

Our definition of neighbors bears some resemblance to the concept of Bounded Neighbors in \cite{KM11},
where the counts of tuples are preserved. As in \cite{KM11}, our
objective is {\it a good choice of neighbors of $D$ (the original
dataset) which should be difficult to distinguish from each other}.
Our differential privacy model retains the essence in
Definition \ref{defn0} from \cite{BLR08}.

\begin{definition}[$\ell'$-diverted privacy]
We say that a non-interactive database sanitization mechanism ${\cal A}$
satisfies $\ell'$-diverted zero-differential privacy, or
simply $\ell'$-diverted privacy, if for any given $D_1$,
for any tuple $t$ in $D_1$,
there exists $\ell' -1$ neighboring databases
$D_2$ with respect to $t$, such that for all sanitized outputs $\hat{D}$,
$Pr[ {\cal A}(D_1) = \hat{D} ] = Pr[ {\cal A}(D_2) = \hat{D} ]$.
\label{defn3}
\end{definition}


The above definition says that any individual may take on any of $\ell'$ different
sensitive values by swapping the sensitive information with other individuals
in the dataset, and it makes no difference in the probability of generating
any dataset $\hat{D}$.
It seems that our definition depends on the parameter $\ell'$.
However, not knowing which $\ell'-1$ neighboring databases it should be
in the definition,
an adversary will not be able to narrow down the possibilities.
Therefore, even in the case where
the adversary knows all the information about all individuals except for 2
individuals, there is still no certainty in the values for the 2
individuals.

Our task is to find a mechanism that satisfies
$\ell'$-diverted zero-differential privacy while at the same time
supports discriminative utilities. The use of Laplace
noise with distribution $Lap(f/\epsilon)$ is common in $\epsilon$-differential \
privacy \cite{D06}. However, this approach will introduce
arbitrary noise when $\epsilon$ becomes zero and it is designed for
interactive query answering. We need to derive
a different technique.

\section{Randomization: An Initial Attempt}

\label{init}

In the search for a technique to guarantee a tapering accuracy
for the estimated values from large counts to small counts,
the law of large numbers \cite{Larson} naturally comes to mind.
Random perturbation has been suggested in \cite{TYW84}, the reason
being that {\it ``If a query set is sufficient large, the law of
large numbers causes the error in the query to be significantly less
than the perturbations of individual records.''}
Indeed, we have seen the use of i.i.d. for the
randomization of datasets with categorical attributes.
In \cite{agrawal05},
an identity perturbation scheme for categorical sensitive values
is proposed. This scheme keeps the original sensitive value
in a tuple
with a probability of $p$ and randomly picks any other
value to substitute for the true value with a probability of $(1-p)$,
with equal probability for each such value.
Theorem 1 in \cite{agrawal05} states that their method can achieve good
estimation for large dataset sizes.
Therefore, it is fair to ask if this approach can
solve our problem at hand. Unfortunately,
as we shall show in the following, this method
cannot guarantee zero-differentiality unless $p$ is equal to $1/M$
with a domain size of $M$, which renders the generated data a totally
random dataset. Let us examine this approach in more details.

Suppose that the tuple $t$ of an individual has sensitive value $t.s$ in $D$.
The set of sensitive values is given by $\{ s_1, ..., s_m \}$.
We generate a sanitized value for the individual by selecting $s_i$ with probability $p_i$,
so that
\[ p_i = \left\{ \begin{array}{ll}
p & \ \mbox{for} \ s_i = t.s;\\
q & \ \mbox{for} \ s_i \neq t.s \end{array} \right.\]
where $\sum_i p_i = 1$.

Let us refer to the anonymization mechanism above by $A$.

Let $D'$ be a dataset published by $A$ 
which contains a tuple for individual $I$.
Consider two datasets $D_1, D_2$ which differ only in the 
sensitive value for the single tuple for $I$.

We are interested in the probability $Pr[A(D_1)=D']$ that $D'$ is 
generated from $D_1$ by $A$,
and $Pr[A(D_2)=D']$.
In particular, we shall show that when $p=q$, $A$ is zero-differential.

Let the tuples in $D_1$ be $t^1_1, ..., t^1_N$.
Let the tuples in $D_2$ be $t^2_1, ..., t^2_N$.

\begin{lemma}
For mechanism $A$, if $p = q$, then $A$ satisfies zero-differential
privacy according to Definition \ref{defn0}, with
neighboring databases
having a symmetric difference of at most 2.
\label{lem1}
\end{lemma}

{\bf Proof}:
Since all the non-sensitive values are preserved and only the sensitive values
may be altered by $A$, we consider the probability that each tuple in $D_1$ or $D_2$
may generate the corresponding sensitive value in $D'$.
For $D_k$, $k \in \{1,2\}$, let $p_k(t_i,s_j)$ be the probability that $A$ will generate $s_j$ for tuple $t_i$.

Mechanism $A$ handles each tuple independently.
Hence $Pr[A(D_k)=D']$ is a function of $p_k(t^k_i,s_j)$ for all $i,j,k \in \{1,2\}$.
\begin{eqnarray*}
Pr[A(D_k)=D'] &=& f(p_k(t^k_1,s_1), p_k(t^k_1,s_2),..., p_k(t^k_1,s_m), \\
& & ..., p_k(t^k_N,s_1),..., p_k(t^k_N,s_m))
\end{eqnarray*}
Given a tuple $t$ with sensitive value $t.s$, the probability that a sensitive
value $s_j$ will be
generated in $D'$ for $t$ depends only on the value of $t.s$.

Without loss of generality, let $D_1$ and $D_2$ differ only in the sensitive value for
$t_r$.
We have $p_1(t^1_i,s_j) = p_2(t^2_i,s_j)$ for all $j$ and all $i \neq r$.
Obviously if we set $p = q = \frac{1}{m}$, then the probability to generate any
value given any original sensitive value will be the same.
Although $t_r^1.s \neq t_r^2.s$, we have
$p_1(t^1_r,s_j) = p_2(t^2_r,s_j)$ for all $j$.
Hence $Pr[A(D_1)=D'] = Pr[A(D_2)=D']$ and Mechanism $A$ satisfies zero-differential
privacy.
$\qed$


\begin{lemma}
For mechanism $A$, if $p \neq q$, then $A$ does not satisfy zero-differential
privacy
according to Definition \ref{defn0}, with
neighboring databases
having a symmetric difference of at most 2.
\end{lemma}

{\bf Proof}:
We prove by constructing a scenario where we are given datasets $D_1, D_2$ differing in
only one tuple, and a sanitized table $D'$, and
$Pr[A(D_1)=D']$ $\neq$ $Pr[A(D_2)=D']$.
Consider the case where all tuples are unique in terms of the
non-sensitive attributes.
For $1 \leq i \leq n$,
let
$p_1(t_i) = p$ if $t_i^1.s = t_i.s$,
and
$p_1(t_i) = q $ if $t_i^1.s \neq t_i.s$.
We have
$Pr[A(D_1)=D'] = \prod_i p_1(t_i)$.
Similarly we define $p_2(t_i)$ for $1 \leq i \leq n$.
$Pr[A(D_2)=D'] = \prod_i p_2(t_i)$.
Furthermore, let $t_k^1.s = t_k.s$ and
$t_k^2.s \neq t_k.s$.
Therefore, for $t_k$, $p_1(t_k) = p$ and $p_2(t_k) = q$,
while $p_1(t_i) = p_2(t_i)$ for $i \neq k$.
\[
\frac{Pr[A(D_1)=D']}{Pr[A(D_2)=D']} = \frac{p}{q}\]
Since $p \neq q$,
it follows that $Pr[A(D_1)=D'] \neq Pr[A(D_2)=D']$ and
therefore $A$ does not satisfy zero-differential privacy.
$\qed$

\begin{lemma}
For mechanism $A$, if $p \neq q$, then $A$ is not
$\ell$'-diverted zero-differential
according to Definition \ref{defn3}.
\end{lemma}
{\bf Proof}:
We proof by showing a scenario where given $D_1$, and a neighboring database
$D_2$ with respect to a tuple $t$, and an anonymized table $D'$, $Pr[A(D_1)=D']$ $\neq$ $Pr[A(D_2)=D']$.
Let $D_1$ and $D_2$ agree on all tuples except for $t_a^1$ and ${t_b^1}$
in $D_1$
and corresponding tuples ${t_a^2}$ and ${t_b^2}$ in $D_2$.
Let all tuples be unique in terms of the
non-sensitive attributes.

For $1 \leq i \leq n$,
let
$p_1(t_i) = p$ if $t_i^1.s = t_i.s$,
and
$p_1(t_i) = q $ if $t_i^1.s \neq t_i.s$.
We have
$Pr[(A(D_1)=D'] = \prod_i p_1(t_i)$.
Similarly we define $p_2(t_i)$ for $1 \leq i \leq n$.
$Pr[A(D_2)=D'] = \prod_i p_2(t_i)$.
Furthermore, let $t_a^1.s = t_a.s$ and
$t_b^1.s = t_b.s$, also
$t_a^1.s \neq t_a.s$ and
$t_b^2.s \neq t_b.s$.
Therefore , for $t_k$, $p_1(t_a) = p$, $p_1(t_a) = p$ and
$p_2(t_a) = q$, $p_2(t_b) = q$,
while $p_1(t_i) = p_2(t_i)$ for $i \not\in \{a,b\}$.
\[
\frac{Pr[A(D_1)=D']}{Pr[A(D_2)=D']} = \frac{p^2}{q^2}\]
Since $p \neq q$,
it follows that $Pr[A(D_1)=D'] \neq Pr[A(D_2)=D']$ and
therefore $A$ is not zero differential.
$\qed$

\medskip

From the previous analysis, in order to make the probability $Pr[A(D_1)=D']$
equal to $Pr[A(D_2)=D']$, all values need to be selected with
probability equal to $\frac{1}{m}$.
This would be the same as random data and
it would have great cost in the utility.

\section{Proposed Mechanism}

From the previous section on mechanism $A$, we see that
for generating a
dataset $D'$ from a given dataset $D$,
randomization with
uniform probability can attain zero-differentiality.
However, if the probability is uniform over the entire domain,
the utility will be very low. Here
we introduce a simple mechanism called $A'$
which introduces uniform probability over a subset of
the domain. We shall show that this mechanism
satisfies $\ell'$-diverted zero-differential privacy
without sacrificing too much utility.
We make the same assumption as in previous works
\cite{l-diversity,XT06b}
that
the dataset is eligible, so that the highest frequency
of any sensitive attribute value does not exceed $N/\ell'$.
Furthermore we assume that $N$ is a multiple of $\ell'$
(it is easy to ensure this by deleting no more than $\ell'-1$
tuples from the dataset).

\subsection{Mechanism $A'$}

Mechanism $A'$ generates a dataset $D'$ given the dataset $D$.
We assume that there is a single sensitive attribute ($SA$) $S$ in $D$.
We shall show that
$A'$ satisfies
$\ell'$-diverted zero-differential privacy.
There are four main steps for $A'$:

\begin{enumerate}
\item

First we assume that the tuples in $D$ have been randomly assigned unique
tuple id's independent of their tuple contents.
Include the tuple id as an attribute $id$ in $D$.
The first step of $A'$ is an initialization step, whereby
the dataset $D$ goes through a projection operation on $id$ and the $SA$
attribute $S$.
Let the resulting table be $D_s$.
That is, $D_s$ $= \Pi_{id,S} ( D )$.
Note that the non-sensitive values have no influence on the generation
of $D_s$.

\item

The set of tuples in $D_s$ is partitioned into sets of size $\ell'$ each
in such a way that in each partition, the sensitive value of each
tuple is unique. In other words,
let there be $r$ partitions, $P_1, ..., P_r$;
in each partition $P_i$, there are $\ell'$ tuples, and
$\ell'$ different sensitive values.
We call each partition a \emph{decoy group}.
If tuple $t$ is in $P_j$, we say that the elements in $P_j$ are the
decoys for $t$. We also refer to $P_j$ as $P(t)$.
With a little abuse of terminology, we also refer to the set of records in $D$
with the same $id$'s as the tuples in this decoy group as $P(t)$.


One can adopt some existing partitioning methods in the literature of
$\ell$-diversity. We require that the method be
deterministic. That is, given a $D_s$, there is a unique partitioning
from this step.
\item

For each given tuple $t$ in $D_s$, we determine the partition $P(t)$.
Let the sensitive values in $P(t)$ be $\{s_1', ..., s_{\ell'}'\}$.
For each of these decoy values, there is a certain probability
that the value is selected for publication as the sensitive
value for $t$.
For a value not in $\{s_1', ..., s_{\ell'}'\}$, the
probability of being published as the value for $t$ is zero.
In the following we shall also refer to the set $\{s_1', ..., s_{\ell'}'\}$
as $decoys(t)$.

Suppose that a tuple $t$ has sensitive value $t.s$ in $D$.
Create tuple $t'$ and initialize it to $t$.
 Next we generate a value to replace the $S$ value in $t'$ by selecting $s_i$ with probability $p_i$,
so that
$$
\begin{array}{lll}
p_i&=p \hspace*{0.8in} &\mbox{for \ }  s_i = t.s \\
p_i&=q =(1-p)\frac{1}{\ell'-1}  &\mbox{for \ } s_i \neq t.s , \  s_i \in decoys(t)\\
p_i&=0  &\mbox{for \ } s_i \not\in decoys(t)\\
\end{array}
$$
\item
The set of tuples $t'$ created in the previous step forms
a table $Ds'$.
Remove the $s$ column from $D$, resulting in $D_N$.
Form a new table $D'$ by joining $Ds'$ and $D_N$ and
retaining only $NSA$ and $S$ in the join result.
The tuples in $D'$ are shuffled randomly. Finally $D'$ and
$\ell'$ are published.
\done
\end{enumerate}

\begin{algorithm}[htb]
\caption{- Mechanism $A'$}
\label{alg:merge}
\begin{algorithmic} [1]
\REQUIRE $D$ with $N$ tuples,with random tuple $id$'s, sensitive attribute $S$, and set of non-sensitive attributes $NSA$
\STATE table $Ds$ $\leftarrow$ $\Pi_{id,S} ( D )$
\STATE partition $D_s$ into decoy groups of size $\ell'$ each\\
\hspace*{9mm} so that each decoy group has $\ell'$ unique sensitive values.
\FOR{each partition $P$}
   \FOR{each tuple $t$ in $P$}
            \STATE let $decoys(t)$ = \{$s'_1, ..., s'_{\ell'}$\}
            \STATE create tuple $t'$ and set $t'.id = t.id$
            \IF {$t.s = s'_i$}
            \STATE
                 set $t'.s = s'_i$ with probability $p$\\
                 set $t'.s$ to $s'_j \neq s'_i$ with probability $q$
             \ENDIF
   \ENDFOR
\ENDFOR
\STATE let $Ds'$ be the set of tuples $t'$ created in the above
\STATE $D'$ $\leftarrow$ $ \Pi_{NSA,S} ( (\Pi_{id,NSA} D) \Join_{id} Ds' )$
\STATE shuffle tuples in $D'$ and publish $D'$ and $\ell'$
\STATE \COMMENT{Note that no other information about the partitions is published }
\end{algorithmic}
\end{algorithm}

The pseudocode for mechanism $A'$ is given in Algorithm 1.
At first glance, mechanism $A'$
looks similar to partitioning based methods for $\ell$-diversity
\cite{l-diversity}, in fact, for the second step in $A'$, we can adopt
an existing partitioning algorithm such as the
 one in \cite{XT06b} which has been designed for bucketization.
However, $A'$ differs from these previous approaches in important ways.

Firstly, the generation of dataset $D'$ is based on a probabilistic
assignment of values to attributes in the tuples. There is a non-zero
probability that an $SA$ (sensitive attribute) value that exists in $D$ does not exist in $D'$.
In known partitioning based methods, the $SA$ values in $D'$
are honest recording of the values in $D$, although in some
algorithms they may be placed in buckets separated from the
remaining values.

Secondly, the partitioning information is not released by $A'$,
in contrast to previous approaches, in which the anonymized
groups or buckets are made known in the data publication.
For $\ell$-diversity methods, since the
partitioning is known, each tuple has a limited set of
$\ell$ possible values.
By withholding the partitioning information, plus the possibility that
a value existing in $D$ may not exist in $D'$, there is essentially
no limit to the possible values for $S$ except for the entire domain
for any tuple in $D'$.

\subsection{$\ell'$-diverted zero-differentiality guaruantee}

For the privacy guarantee, we shall show that if
$p = q$, then $A'$ satisfies $\ell$'-diverted zero-differential
privacy, otherwise, it does not.
First we need to state a fact about $A'$.

\begin{fact}
In mechanism $A'$, let $p = q = \frac{1-p}{\ell' - 1}$, so that $p =  q = \frac{1}{\ell'}$.
When executing $A'$, for two tuples $t$ and $t'$ in the same
partition ($P(t)=P(t')$), and any sensitive value $s_i$,
the probability that $t$ will be assigned $s_i$ by $A'$ is
equal to that for $t'$.
\label{guar}
\end{fact}

The following theorem says that we should set $p = q = 1/\ell'$
in mechanism $A'$.

\begin{theorem}
For mechanism $A'$, if $p = q = \frac{1}{\ell'}$,
then $A'$ satisfies $\ell'$-diverted zero-differential privacy.
\end{theorem}

{\bf Proof}:
Let $D'$ be a published dataset.
Given a dataset $D_1$ which may generate $D'$, and a tuple $t$ in $D_1$,
we find $\ell' - 1$ neighboring databases $D_2$ as follows:

We execute $A'$ on top of $D_1$.
In the first step, $D^1_s$ is generated from $D_1$.
In the second step, $D^1_s$ is partitioned into sets of size $\ell'$.
Let $P(t)$ be the partition (decoy group) formed by $A'$ for $t$ in $D_1$.
Pick one element $\breve{t}$ in $P(t)$ where $\breve{t} \neq t$.
Form $D_2$ by
swapping the non-sensitive values of $t$ and $\breve{t}$ in $D_1$.
By definition $D_2$ is a neighboring database of $D_1$.

Let $D^2_s$ be the table generated from Step 1 of $A'$ on $D_2$.
Since we have only swapped the non-sensitive values of $t$ and $\breve{t}$,
$D^1_s = D^2_s$.
The partitioning step of $A'$ is deterministic, meaning that the same partitioning
will be obtained for $D_1$ and $D_2$.
From the above, we know that $t$ and $\breve{t}$ are in the same partition for $D_1$,
i.e.,
$P(t) = P(\breve{t})$.
When we consider the generation of
sensitive values for $t$ and $\breve{t}$, since they are in the same
partition $P(t)$, by Fact \ref{guar}, they have the same probabilities
for different outcomes.
Since the $SA$ values for different
tuples are generated independently, $Pr[A'(D_1)=D'] = Pr[A'(D_2)=D']$.

There are $\ell'-1$ possible $D_2$ given $D_1$,
we have shown that $A'$
satisfies $\ell'$-diverted zero-differential privacy.
\done

\begin{theorem}
For mechanism $A'$, if $p \neq q$, then $A'$ does not satisfy $\ell'$-diverted
zero-differential privacy.
\end{theorem}

{\bf Proof}:
We prove by giving an instance where $A'$ is not
$\ell'$-diverted zero-differential.
We say that a dataset $D$ is $A'$-consistent with $D'$ if
there is a non-zero probability that $D'$ is generated
from $D$ by $A'$.
Consider $D_1$ and $D_2$, each being consistent with $D'$.
Let the tuples in $D_1$ be $t^1_1, ..., t^1_N$.
Let the tuples in $D_2$ be $t^2_1, ..., t^2_N$.
The two sets of tuples are for the same set of individuals.
Furthermore, assume that $D_1$ and $D_2$
differ in only 2 tuples for a pair of individuals;
let the pair of tuples in $D_1$ be $t_a^1$, $t_b^1$,
and that in $D_2$ be $t_a^2$, $t_b^2$. Assume that all
tuples have unique non-sensitive values, and
\begin{eqnarray*}
t_a^1.s = t'_a.s,
t_b^1.s = t'_b.s\\
t_a^2.s \neq t'_a.s,
t_b^2.s \neq t'_b.s\\
t_a^1.s = t_b^2.s,
t_a^2.s = t_b^1,s
\end{eqnarray*}
For $1 \leq i \leq N$, let
$p_1(t_i) = p$ if $t_i^1.s = t_i'.s$, and
$p_1(t_i) = q $ if $t_i^1.s \neq t_i'.s$.
Similarly we define $p_2(t_i)$ for $1 \leq i \leq N$.

Therefore , for $t_a$, $p_1(t_a) = p_1(t_b) = p$ and $p_2(t_a) = p_2(t_b) = q$,
while $p_1(t_i) = p_2(t_i)$ for $i \not\in \{a,b\}$.
\[
\frac{Pr[A'(D_1)=D']}{Pr[A'(D_2)=D']} = \frac{\prod_i p_1(t_i)}{\prod_i p_2(t_i)} = \frac{p^2}{q^2} \]
Since $p \neq q$,
it follows that $Pr[A'(D_1)=D'] \neq Pr[A'(D_2)=D']$ and
therefore $A$ is not zero differential.
\done

\medskip

The above theorems show that in order to enforce $\ell'$-diverted zero-differential privacy, we should set both $p$ and $q$ to $1/\ell'$.
This will be the assumption in our remaining discussions about
$A'$. 

\section{Aggregate Estimation}
\label{secUtilization}

In this section we examine how to answer counting queries
for the sensitive attribute based on the published dataset $D'$.

Let $|D| = N$, so that there are $N$ tuples in $D$.
Consider a sensitive value $s$.
Let the true frequency of $s$ in $D$ be $f_s$. By mechanism $A'$,
there will be $f_s$ decoy groups which contain $s$ in the decoy
value sets.  Each tuple in these groups has a probability of
$p = \frac{1}{\ell'}$ to be assigned $s$ in $D'$.
The probability that it is assigned other values $\bar{s}$ is $1-p$.
There are $f_s \ell'$ such tuples.

Let $N'_s$ denote the number of times that $s$ is published in $D'$.
The random variable $N'_s$ has the binomial distribution with
parameters $f_s \ell'$ and $p$.
$$P\left[ N'_s = x \right] = {{f_s \ell'}\choose x}p^{x} (1-p)^{f_s \ell' -x}$$
The expected value is $f_s \ell' p$, and
$\sigma^2 = f_s \ell' p (1-p)$

Since we set $p=q = 1/\ell'$, the expected count of $s$ in $D'$ is given by
$e_s = p \ \ell' f_s  = f_s$, we have
\begin{eqnarray*}
e_s = f_s
\end{eqnarray*}
 That is, to estimate the true count of an $SA$ value $s$, we simply take the count of $s$ in $D'$, $f'_s$.

\begin{theorem}
The estimation of $f_s$ by $f'_s$ is a maximum likelihood estimation (MLE).
\label{lemML}
\end{theorem}

{\bf Proof}.
Let $L(D)$ be the likelihood of the observation $f'_s$ in $D'$,
given the original dataset $D$.
$L(D) = Pr(f'_s|D)$

From Mechanism A', given $f_s$ occurrences of $s$ in $D$, there will
be exactly $\ell' f_s$ tuples that generates $s$ in $D'$ with a
probability of $p$. The remaining tuples have zero probability of
generating a $s$ value.
The probability that $f'_s$ occurrences of $s$ is generated in $D'$
is given by
$$
L(D) = Pr(f'_s|D) = {{\ell' f_s} \choose f'_s} p^{f'_s}  (1-p)^{\ell' f_s - f'_s}
$$
where $p = 1/\ell'$.

This is a binomial distribution function which is maximized
when $f'_s$ is at the
mean value of $\ell'f_s p = f_s$.
\done

\medskip

To examine the utility of the dataset $D'$, we ask
how likely $f'_s$ is close to $f_s$,
and hence the estimation $e_s$ is close to the true count $f_s$?
However, we also need to provide protection for small counts.
In the next section we shall analysis these
properties of the published dataset.

\section{Privacy, Utility, and the sum}
\label{sec:sec6}

As discussed in Section \ref{issues}, the utility of the dataset
must be bounded so that for certain facts, in particular, those
that involve very few individuals, the published data should
provide sufficient protection. Here we consider the relationship between
the utility and the number of tuples $n$ that is related
to a sensitive value.
Is it possible to balance between
disclosing useful information where $n$ is large and hence safe
and not disclosing accurate information when $n$ is small and hence need
protection? We explore these issues in the following.

\subsection{Utility for large sums}

To answer the question about the utility for
large sums, we make use of the Chebychev's inequality
which gives a bound for the likelihood that an observed value deviates
from its mean.

\medskip

{\bf Chebychev's Theorem}: If $X$ is a random variable with mean $\mu$
and standard deviation $\sigma$, then for any positive $k$,
$ Pr ( |X - \mu| < k \sigma ) \geq 1 - \frac{1}{k^2} $
and
$ Pr ( |X - \mu| \geq k \sigma ) \leq \frac{1}{k^2} $
\done

\bigskip

Let $X_1, X_2, ..., X_n, ...$ be a sequence of
independent, identically distributed random variables, each
with mean $\mu$ and variance $\sigma^2$.
Define the new sequence of $\overline{X}_i$ values by
$$\overline{X}_n = \frac{1}{n} \sum_{i=1}^n X_i,  n = 1, 2, 3, ...$$

From Chebychev's inequality,
$P\left[| \overline{X}_n - \mu_{\overline{X}_n} | \geq k
\sigma_{\overline{X}_n} \right] \leq \frac{1}{k^2}$
where
$\mu_{\overline{X}_n} = E[\overline{X}_n] = \mu$,
$\sigma_{\overline{X}_n} = E[( \overline{X}_n - \mu)^2] = \frac{\sigma^2}{n}$
and
$k$ is any positive real number.
Choose $k = \frac{\epsilon \sqrt{n}}{\sigma}$ for some $\epsilon > 0$,
we get
\begin{eqnarray}
Pr\left[ | \overline{X}_n - \mu | \geq \epsilon \right] \leq \frac{\sigma^2}{\epsilon^2 n}
\label{eqn1}
\end{eqnarray}

The above reasoning has been used to prove the law of large numbers.
Let us see how it can help us to derive the utility of our published
data for large sums. If there are $f_s$ tuples with $s$ value, then $n = \ell' f_s$ tuples in $D$ will have a probability of $p$
to be assigned $s$ in $D'$.
The setting of value $s$ to the tuples in $D'$ corresponds to
a sequence of $\ell' f_s$ independent Bernoulli
random variables, $X_1, ..,, X_{\ell' f_s}$, each with parameter $p$.
Here $X_i = 1$ corresponds to the event that $s$ is chosen for the $i$-th tuple, while $X_i = 0$ corresponds to the case where $s$ is not chosen.

The mean value  $\mu_{\overline{X}_n} = p$.
Also, $\sigma_{\overline{X}_n}^2 = p(1-p)/n$. From Inequality (\ref{eqn1}),
\begin{eqnarray*}
Pr\left[ | \overline{X}_n - \mu | \geq \epsilon \right] \leq \frac{p(1-p)}{\epsilon^2 n^2}
= \frac{p(1-p)}{\epsilon^2 \ell'^2 f_s^2}
\end{eqnarray*}
From Section \ref{nondiff}, we set $p = \frac{1}{\ell'}$, hence
\begin{eqnarray}
Pr\left[ | \overline{X}_n - \mu | \geq \epsilon \right] \leq \frac{1}{\ell'^3\epsilon^2 f_s^2}
\label{eqn3}
\end{eqnarray}

Note that $\overline{X}_n$ is the count of $s$ in $D'$ divided by $n$, and
$n = \ell' f_s$.
Hence the occurrence of $s$ in $D'$ is  $f_s' = \ell' f_s \overline{X}_n$.
Rewriting Inequality (\ref{eqn3}), we get

\begin{eqnarray*}
Pr\left[ | \ell' f_s \overline{X}_n - \ell' f_s \mu | \geq \ell' f_s \epsilon \right] \leq \frac{1}{\ell'^3\epsilon^2 f_s^2}
\end{eqnarray*}
Since $\mu = p = 1/\ell'$,
$Pr\left[ |  f_s' -  f_s | \geq \ell' \epsilon f_s \right] \leq \frac{1}{\ell'^3\epsilon^2 f_s^2}$

With the above inequality, we are interested in how different $f_s'$ is from $f_s$.
Since the deviation is bounded by $\ell' \epsilon f_s$, it is better to
use another variable $\varepsilon = \ell' \epsilon$ to quantify the difference.
\begin{eqnarray}
Pr\left[ |  f_s' -  f_s | \geq  \varepsilon f_s \right] \leq \frac{1}{\ell'\varepsilon^2 f_s^2}
\label{eqn4}
\end{eqnarray}
Our estimation is $e_s = f'_s$, hence the above gives a bound on the
probability of error in our estimation.
If $f_s$ is small, then the bound is large.
In other words the utility is not guaranteed. This is our desired effect.

Given a desired $\varepsilon$ and a desired $\ell'$, we may find a
frequency threshold
${\cal_T}_f$ so that
for $f_s$ above this threshold, the probability of error in
Inequality (\ref{eqn4}) is below another threshold
${\cal T}_E$ for utility.
We can set the RHS in the above
inequality to be this threshold. Obviously,
$\frac{1}{\ell'\varepsilon^2 f_s^2} \leq {\cal T}_E \mbox{ \ for \ }  f_s \geq {\cal T}_f$

\begin{definition}[Thresholds ${\cal T}_E$ and ${\cal T}_f$]
Given an original dataset $D$ and an anonymized dataset $D'$.
A value $s$ has a ($\epsilon$, ${\cal T}_E$,${\cal T}_f$) utility guarantee if
for a frequency $f_s$ of $s$ above the frequency threshold of ${\cal T}_f$ in $D$,
\begin{eqnarray}
Pr\left[ |  f_s' -  f_s | \geq \varepsilon f_s \right] \leq {\cal T}_E
\mbox{ for } f_s \geq {\cal T}_f
\label{eqn5}
\end{eqnarray}
\end{definition}

The above definition says that a value
$s$ has a ($\epsilon$, ${\cal T}_E$,${\cal T}_f$) guarantee if
 whenever the frequency $f_s$ of $s$ is above
${\cal T}_f$ in $D$, then
the probability of a relative error of more than $\varepsilon$ is at most ${\cal T}_E$.

\begin{lemma}
Mechanism $A'$
provides a ($\epsilon$, ${\cal T}_E$,${\cal T}_f$) utility guarantee for each
sensitive value, where
\begin{eqnarray}
\frac{1}{\ell' \varepsilon^2 {\cal T}_E } = {\cal T}_f^2
\end{eqnarray}
\end{lemma}

Hence given, $\varepsilon$ and ${\cal T}_E$,
we can determine the smallest count which can provide
the utility guarantee.

\begin{example}
Consider some possible values for the parameters.
Suppose ${\cal T}_E = 0.02$ and $\ell' = 10$.
If $\varepsilon = 0.2$, then ${\cal T}_f = 11$.
If $\epsilon = 0.001$,
or $\varepsilon = 0.02$
then ${\cal T}_f = 49$.\done
\end{example}

Figure \ref{fig8} shows the relationship between the possible
values of ${\cal T}_f$ and ${\cal T}_E$.
The utility is better for small ${\cal T}_E$, and the value of ${\cal T}_E$
becomes very small when the count is increasing towards 900.
Note that utility is the other side of privacy breach, it also
means that for concepts with large counts, privacy
protection is not guaranteed, since the accuracy in the count
will be high.

%

\begin{figure}[tbp]
\begin{center}
\includegraphics[width = 0.8\linewidth]{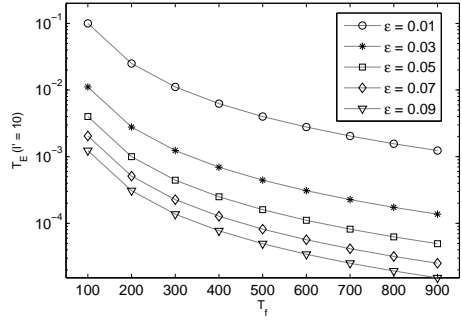}
\end{center}
\vspace{-6mm}
\caption{Relationship between ${\cal T}_E$ and ${\cal T}_f$}
\label{fig8}
\end{figure}

\subsection{Privacy for small sums }

\label{moreProtect}

Next we show how our mechanism can inherently provide protection for
small counts. From Inequality (\ref{eqn4}),
small values of $f_s$ will weaken the guarantee of utility.
We can in fact give a probability for relative errors based on
the following analysis.

The number of $s$ in $D'$ is the total number of successes in $f_s \ell'$ repeated
independent Bernoulli trials with probability $\frac{1}{\ell'}$ of
success on a given trial.
It is the binomial random variable with parameters $n = \ell'f_s$
and $p = \frac{1}{\ell'}$.
The probability that this number is $x$ is given by
$$
{{n} \choose x} p^x  q^{n-x}
=
{{\ell' f_s} \choose x} \left(\frac{1}{\ell'}\right)^x \left(1 - \frac{1}{\ell'}\right)^{\ell'f_s - x}
$$

\begin{example}
If $f_s = 5$, $\ell'=10$, for an $\varepsilon = 0.3$ bound on
the relative error, we are interested to know how
likely $f'_s$ is close to 5 within a deviation of 1. The
probability that $f'_s$ is between 4 to 6 is given by
\begin{eqnarray*}
\sum_{x = 4}^{6} {{n} \choose x} p^x  q^{n-x}
=
\sum_{x = 4}^{6} { 50 \choose x } 0.1^x 0.9^{50-x}
\approx 0.52
\end{eqnarray*}
Hence the probabilty that $f'_s$ deviates from $f_s$ by more than $0.3 f_s$
is about 0.52.
\label{egsmall}
\end{example}

\begin{definition}[privacy guarantee]
We say that a sensitive value $s$ 
has a $(\varepsilon, {\cal T}_P)$ privacy guarantee if
the probability that the estimated count of $s$, $f'_s$, has a
relative error of more than $\varepsilon$
is at least ${\cal T}_P$.
\end{definition}

In Example \ref{egsmall}, the value $s$ has a $(0.3, 0.48 )$ privacy guarantee.
A graph is plotted in Figure \ref{fig62}
for the expected error for small values of $f_s$.
Here
the summation in the above probability
is taken from $f'_s = \lceil 0.7 f_s \rceil$ to $f'_s = \lfloor 1.3 f_s \rfloor$.
We have plotted for different $f_s$ values the probability given by
\begin{eqnarray*}
1 - \sum_{x = \lceil 0.7 f_s \rceil}^{\lfloor 1.3 f_s \rfloor} {{n} \choose x} p^x  q^{n-x}
\end{eqnarray*}
This graph shows that the relative error in
the count estimation is expected to be large for sensitive values with small counts.

\begin{figure}[tbp]
\begin{center}
\includegraphics[width = 0.8\linewidth,height=1.8in]{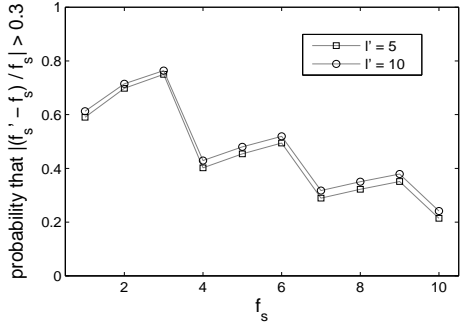}
\end{center}
\vspace{-6mm}
\caption{Expected error for small sums}
\label{fig62}
\end{figure}

\section{Multiple Attribute Predicates}
\label{secCorr}

In this section we consider the counts for sets of values.
For example, we may want to know the count of tuples
with both lung cancer and smoking, or the count of
tuples with gender = female, age = 60 and disease = allergy.
The problem here is counting the occurrences of values of an attribute set.
Firstly we shall consider counts for predicates
involving a single sensitive attribute, then we extend
our discussion to predicates
involving multiple sensitive attributes.

\subsection{Predicates involving a single $SA$}

Assume that we have a set of non-sensitive attributes $NSA$ and
a single sensitive attribute $SA$, let us consider queries involving
both $NSA$ and $SA$.
We may divide such a query into two components: $P$ and $s$,
where $P \in domain(NA)$ ($NA \subseteq NSA$), and $s \in domain(SA)$.
For example $P = (female, 60)$ and $s = (allergy)$.
Note that the non-sensitive attributes
are not distorted in the published dataset.
This can be seen as a special case of
generating a non-sensitive value for the individual $t$ by
selecting $s_i$ with probability $p_i$,
so that

\[ p_i = \left\{ \begin{array}{ll}
1 & \ \mbox{for} \ s_i = t.s;\\
0 & \ \mbox{for} \ s_i \neq t.s. \end{array} \right.\]

Suppose we are interested in the count of the co-occurrences of
non-sensitive values $P$ and $SA$ $s$.

\begin{definition}[state $i$]
There are 4 conjunctive predicates concerning $P$ and $s$,
namely, $p_0 = \overline{P} \wedge \overline{s}$,
$p_1 = \overline{P} \wedge {s}$,
$p_2 = {P} \wedge \overline{s}$, and
$p_3 = {P} \wedge {s}$.
If a tuple satisfies $p_i$, we say that it is at state $i$.
\end{definition}

The distributions of the predicates in $D$ and $D'$ are
given by $cnt(p_i)$ and $cnt'(p_i)$, respectively.
Here $cnt(p_i)$($cnt'(p_i)$) is the number of tuples satisfying $p_i$
in $D$ ($D'$).

For simplicity we let $x_i = cnt(p_i)$ and $y_i = cnt'(p_i)$,
hence the a priori distribution concerning the states in $D$ is given by $x = \{x_0, x_1, x_2, x_3\}$,
and the distribution in $D'$ is given by $y = \{y_0, y_1, y_2, y_3\}$.
Hence $y$ contains the observed frequencies.

\begin{definition}[Transition Matrix $M$]
The probability of transition for a tuple from an initial state $i$ in $D$ to
a state $j$ in $D'$ is given by $a_{ij}$.
The values $a_{ji}$ forms a transition matrix $M$.
\end{definition}

The values of $a_{ij}$ are given in Figure \ref{figa}.

\begin{figure}[h]
\begin{small}\begin{center}
\begin{tabular}{ c | c c c c }
 & \ $y_0$($\overline{P}\overline{s}$) \ & \ $y_1$($\overline{P}s$) \ & \ $y_2$($P \overline{s}$) \ & \
        $y_3$($ P s $)\ \\
        &&&&\\
        \cline{2-5}\\
 &  $a_{00}=$ & $a_{01}=$ & $a_{02}=0$ & $a_{03}=0$ \\
  $x_0$($\overline{P}\overline{s}$)& $ 1 - a_{01}$  & $\frac{x_1+x_3}{N}$ &   &   \\
  &&&&\\
  & $a_{10}= $ & $a_{11}= $  & $a_{12}=0$ & $a_{13}=0$ \\
  $x_1$($\overline{P} s$)
  & $\frac{\ell'-1}{\ell'}$ & $\frac{1}{\ell'}$ &&\\
  &&&&\\
 &  $a_{20}=0$ & $a_{21}=0$  & $a_{22}=$ & $a_{23}=$ \\
  $x_2$($P \overline{s}$)  &  & &  $1 - a_{23}$ & $\frac{x_1+x_3}{N}$\\
   &&&&\\
 & $a_{30}=0$  &  $a_{31}=0$ &  $a_{32}=$ & $a_{33}=$ \\
  $x_3$($P s $) &&&$\frac{\ell'-1}{\ell'}$&$\frac{1}{\ell'}$\\
  &&&&\\
  \hline
\end{tabular}
\caption{State transition probabilities}
\label{figa}
\end{center}\end{small}
\end{figure}

Let $Pr(r_i|x)$ be the probability that a tuple has state $i$ in $D'$
given vector $x$ for the initial state distribution.
The following can be derived.
\begin{eqnarray}
Pr(r_0|x) &=& \frac{1}{N} \left( \left(1 - \frac{x_1+x_3}{N}\right) x_0 + \frac{\ell' - 1}{\ell'} x_1  \right) \label{eq1}\\
Pr(r_1|x) &=& \frac{1}{N} \left( \left(\frac{x_1+x_3}{N}\right) x_0 + \frac{1}{\ell'} x_1  \right)\\
Pr(r_2|x) &=& \frac{1}{N} \left( \left(1 - \frac{x_1+x_3}{N}\right) x_2 + \frac{\ell'-1}{\ell'} x_3  \right)\\
Pr(r_3|x) &=& \frac{1}{N} \left( \left(\frac{x_1+x_3}{N}\right) x_2 + \frac{1}{\ell'} x_3  \right)
\label{eq4}
\end{eqnarray}
The above equations are based on the mechanism generating $D'$ from $D$.
Let us consider the last equation, the other equations are derived in
a similar manner.
For each true occurrence of $(P, s)$, there is
a $\frac{1}{\ell'}$ probability that it will generate such an occurrence
in $D'$.
If there are $x_3$ such tuples, then the expected number of
generated instances will be
${x_3}/{\ell'}$.

Other occurrences of $(P,s)$ in $D'$ may be generated by the
$x_2$
tuples satisfying $P$ but with $t.s \neq s$ $(P,\overline{s})$. Each such
tuple $t$ satisfies $P$ for the non sensitive values and it is possible that
$s \in decoys(t)$.
We are interested to know how likely $s \in decoys(t)$.

There are in total $\frac{N}{\ell'}$ partitions. There can be
at most one $s$ tuple in each partition. Hence
$f_s$ of the partitions contain $s$ in the decoy set,
and if a tuple $t$ is in such a partition, then $s \in decoys(t)$.
The probability
of having $s$ in $decoys(t)$ for a tuple $t$ with $t.s \neq s$
is the probability that $t$ is in one of the $f_s$ partitions above.
Since mechanism $A'$ does not consider
the $NSA$ values in the randomization process, all such tuples $t$
have equal probability of being in any of the $f_s$ partitions,
and the probability is
given by $f_s/\frac{N}{\ell'} = f_s\frac{\ell'}{N}$. Since $f_s = x_1 + x_3$,
this probability is $\frac{x_1 + x_3}{N} \ell'$.

The total expected occurrence of $(P,s)$ is given by
\begin{eqnarray*}
 \frac{x_3}{\ell'} + \left( \frac{x_1+x_3}{N} \ell'\right) \frac{x_2}{\ell'}
\end{eqnarray*}

We can convert this into a conditional probability that a tuple in $D'$ satisfies
$(P,s)$ given $x$, denoted by $Pr(r_3|x)$. This gives Equation (\ref{eq4}).

Rewriting Equations (\ref{eq1}) to (\ref{eq4}) with the transition
probabilities in Figure \ref{figa} gives the following:
\begin{eqnarray}
\label{eqcorr}
Pr(r_i|x) = \sum_{j=0}^3 a_{ji}  \frac{x_j}{N}
\end{eqnarray}

Equation (\ref{eqcorr}) shows that $a_{ji}$ is the probability
of transition for a tuple from an initial state $j$ in $D$ to
a state $i$ in $D'$.
%

We adopt the iterative Bayesian technique for the estimation
of the counts of $x_0, ..., x_3$. This method is similar to the
technique in \cite{agrawal05} for reconstructing multiple column
aggregates.

Let the original states of tuples $t_1, ..., t_N$ in $D$ be
$U_1, ..., U_N$, respectively.
Let the states of the corresponding tuples in $D'$ be
$V_1, ..., V_N$.
From Bayes rule, we have
\begin{eqnarray*}
Pr(U_k = i| V_k = j) = \frac{P(V_k = j | U_k = i)P(U_k = i)}{P(V_k = j)}
\end{eqnarray*}
Since $Pr(U_k = i) = x_i/N$, and $Pr(V_k = j| U_k = i) = a_{ij}$,
\begin{eqnarray}
Pr(U_k = i|V_k = j)
= \frac{a_{ij} \frac{x_i}{N}}{\sum_{r=0}^3 a_{rj}\frac{x_r}{N}}
\label{eqna}
\end{eqnarray}
\begin{eqnarray*}
Pr(U_k = i) = \sum_{j=0}^3 Pr(V_k = j) Pr(U_k = i|V_k = j)
\end{eqnarray*}
Hence, since $Pr(V_k = j) = y_j/N$, $Pr(U_k = j) = x_j/N$ and
from Equation (\ref{eqna}), we have
\begin{eqnarray*}
\frac{x_i}{N} = \sum_{j=0}^3 \frac{y_j}{N} \frac{a_{ij} \frac{x_i}{N}}{\sum_{r=0}^3 a_{rj}\frac{x_r}{N}}
\end{eqnarray*}
We iteratively update $x$ by the following equation
\begin{eqnarray}
x_i^{t+1} = \sum_{j=0}^3 y_j \frac{a^t_{ij} x^t_i}{\sum_{r=0}^3 a^t_{rj} x^t_r}
\label{eqnb}
\end{eqnarray}

We initialize $x^0 = y$, and $x^t$ is the value of $x$ at iteration $t$.
In Equation (\ref{eqnb}), $a^t_{ij}$ refer to the value of $a_{ij}$ at
iteration $t$, meaning that the value of $a^t_{ij}$ depends on
setting the values of
$x=x^t$. We iterate until $x^{t+1}$ does not differ much from $x^t$.
The value of $x$ at this fixed point is taken as the estimated $x$ values.
In particular $x_3$ is the estimated count of $(P,s)$.

For the multiple attribute predicate counts, we also guarantee that
privacy for small sums will not be jeopardized.

\begin{lemma}
Let $s$ be a sensitive value with a $(\varepsilon, {\cal T}_p)$ privacy guarantee,
then the count for a multiple column aggregate
involving $s$ also has the same privacy guarantee.
\end{lemma}

{\bf Proof}:
Without loss of generality, consider a multiple attribute aggregate
of $(P,s)$, where $P \in domain(NSA)$. Since the randomization of $s$ is independent of the
$NSA$
attributes, the expected relative error
introduced for $(\overline{P},s)$ is the same as that for $(P,s)$.
The total expected error for $(\overline{P},s)$ and $(P,s)$ must not be less
than that dictated by the $(\varepsilon, {\cal T}_P)$ guarantee
since otherwise the sum of the two counts
will generate a better estimate for the count of $s$, violating
the $(\varepsilon, {\cal T}_P)$ privacy for $s$.
Hence for $(P,s)$ the privacy guarantee is
at least $(\varepsilon, {\cal T}_P)$.
\done

\label{sec7}

\subsection{Multiple sensitive attributes}

So far we have considered that there is a single sensitive attribute in
the given dataset.
Suppose instead of a single sensitive attribute ($SA$), there are multiple $SA$s, let the sensitive attributes be $S_1, S_2, ... S_w$.
We can generalize
the randomization process by treating each $SA$ independently,
building decoy sets for each $S_i$.

For predicates involving $\{P, s_1, s_2, ..., s_w\}$, where $P$
is a set of values for a set of non-sensitive attributes, $s_i \in domain(S_i)$,
there will be $K = 2^{w+1}$ different possible states for each tuple.
We let $(P, s_1, s_2, ..., s_w)$ stand for
$(P \wedge s_1 \wedge s_2  ...\wedge s_w)$.
For reconstruction of the count for $(P, s_1, s_2, ..., s_w)$,
we form a transition matrix for all the $K=2^{w+1}$ possible states.
It is easy to see that the case of a single SA in Section \ref{sec7} is a special case
where the transition matrix $M$ is the tensor product of two matrices
$M_0$ and $M_1$, $A = M_0 \bigotimes M_1$, where $M_0$ is for the set of non-sensitive values
and $M_1$ is for $s_1$, and they are defined as follows:

\[ M_0 = \left( \begin{array}{cc}
1 & 0 \\
0 & 1  \end{array} \right) \hspace*{5mm}
M_i = \left( \begin{array}{cc}
1-\frac{f_{s_i}}{N} & \frac{f_{s_i}}{N} \\
\frac{\ell'-1}{\ell'} & \frac{1}{\ell'} \end{array} \right)\]

In general, with sensitive attributes $S_1, ..., S_w$,
the transition matrix is given by $M = M_0 \bigotimes M_1 ... \bigotimes M_w$.

Let the entries in matrix $M$ be given by $m_{ij}$.
We initialize $x^0 = y$ and iteratively update $x$ by the following equation
\begin{eqnarray}
x_i^{t+1} = \sum_{j=0}^{K-1} y_j \frac{m^t_{ij} x^t_i}{\sum_{r=0}^{K-1} m^t_{rj} x^t_r}
\label{eqnc}
\end{eqnarray}

In Equation (\ref{eqnc}), $x^t$ is the value of $x$ at iteration $t$.
$a^t_{ij}$ refer to the value of $m_{ij}$ at
iteration $t$, meaning that the value of $m^t_{ij}$ depends on
setting the values of
$x=x^t$. We iterate until $x^{t+1}$ does not differ much from $x^t$.
The value of $x$ at this fixed point is taken as the estimated $x$ values.
In particular $x_{K-1}$ is the estimated count of $(P,s_1,...,s_w)$.

\section{Belief about an Individual}

An adversary may be armed with auxiliary knowledge
in the attack on the sensitive value of an individual.
In general auxiliary knowledge allows an adversary to rule out possibilities
and sharpen their belief about the sensitive value of
an individual.
For example, a linkage attack refers to an attack with the
help of knowledge about another database which is linked
to the published data. The other database could be a voter
registration list, and it has been discovered that only the
values of birthdate, sex and zip code are often sufficient to identify
an individual
\cite{Sweeney97,samarati-protecting}.

In the design of $\ell$-diversity \cite{l-diversity}, the set of tuples are
divided into blocks and there should be $\ell$ {\it well represented}
sensitive values in each block. The adversary needs $\ell -1$
damaging pieces of auxiliary knowledge to eliminate $\ell -1$
possible sensitive values and uncover the
private information of an individual. Our method is an improvement
over the $\ell$-diversity model since the possible sensitive values
in our case is the entire domain of the sensitive attribute, including
values that do not appear in the dataset. Hence if the domain
size is $m$, the adversary would need $m -1$ pieces of auxiliary
knowledge to rule out $m-1$ possible values, but in that case,
the adversary knows a priori the exact value without examining $D'$.

Another form of auxiliary knowledge is knowledge about
the sanitization mechanism. Since many known approaches aim
to minimize the distortion to the data, they
suffer from minimality attack \cite{WFW+07}.
Our method does not involve any distortion minimization step
and therefore minimality attack will not be applicable. 

%

\section{Empirical Study}

We have implemented our mechanism $A'$ and compared with some existing
techniques that are related in some way to our method.

For step 2 of mechanism $A'$, we need to partition tuples in $D_s$
into sets of size $\ell'$ each and each partition contains $\ell'$
different sensitive values.
We have adopted the group creation step in the algorithm
for Anatomy \cite{XT06b}. In this algorithm, all tuples of the given table are
hashed into buckets by the sensitive values, so that each bucket
contains tuples with the same $SA$ value.
The group creation step consists of multiple iterations. In each
iteration a partition (group) with $\ell'$ tuples is created.
Each iteration has two sub-steps: (1) find the set $L$ with the
$\ell'$ hash buckets that currently have the largest number of tuples.
(2) From each bucket in $L$, randomly select a tuple to be included in
the newly formed partition.
Note that the random selection in step (2) can be made
deterministic by picking the tuple with the smallest
tuple id.

\subsection{Experimental setup}

The experiments evaluate both effectiveness and efficiency of 
mechanism $A'$ for $\ell'$-diverted
privacy. We also compare our method with three other approaches, Anatomy
for $\ell$-diversity,
differential privacy by means of Laplacian perturbation,
and global randomization (mechanism $A$). Our code is written in C++ and executed on a PC with
CORE(TM) i3 3.10 GHz CPU and 4.0 GB RAM. The dataset is generated by randomly
sampling 500k tuples from the
CENSUS\footnote{Downloadable at
http://www.ipums.org} dataset which contains the information
for American adults. We further produce five datasets from the 500k dataset,
with cardinalities ranging from 100k to 500k. The default cardinality is 100k.
Occupation is chosen as the sensitive
attribute, which involves 50 distinct values.

In the experiment we consider count queries, which have been used for
utility studies for partition-based methods \cite{XT06b} and randomization-based
methods \cite{RSH07}. A pool of 5000 count queries is generated according to the method
described in Appendix 10.9 in \cite{CW10}.
Specifically, we generate random predicates on the non-sensitive attributes, each of which
is combined with each of the values in the domain of the sensitive attribute to form a query.
We count the tuples satisfying
a condition of the form $A_1=v_1 \wedge ... \wedge A_d = v_d \wedge SA = v_s$, where each $A_i$
is a distinct non-sensitive attribute, $SA$ is the sensitive attribute, and the $v_i$ and $v_s$
are values from the domains of $A_i$ and $SA$, respectively.
The selectivity of a query is defined as the percentage of tuples that
satisfy the conditions in the query. For each selectivity $s$ that is considered
we report on the average relative error of the estimated count for all queries
that pass the selectivity threshold $s$.
In later analysis, we group queries according to their distinct selectivities.

Given queries in the pool, we calculate the average relative error between the actual
count (from the original dataset)  and estimated count (from the published dataset)
as the metric for utility. As discussed earlier, we differentiate between small counts
and large counts. Specifically, we vary the selectivity (denoted by $s$, which is the ratio
of the actual count to the cardinality of dataset) from $0.5\%$ up to $5\%$ for
large counts. For small counts, we require the actual count to be no more than $10$ (selectivity
less than $0.1\%$). We evaluate the influence of various $\ell$' values, and also the
cardinalities of dataset on the utility. To assess the efficiency, we record and show the running time of
our data publishing algorithm.

\subsection{Utility for large counts}

First we shall examine the impact of varying $\ell$', while we have separate plot
for distinct selectivity. In particular, the average relative error is computed for
$\ell$' ranging from $2$ to $10$, as shown in Figure \ref{fig11} where selectivity of large
counts is concerned. For large selectivity (i.e., large counts) between 2\%
and 5\%, the error is as low as 20\%. The error is also bounded by 40\% for
other selectivities, which is acceptable. Another observation is a trend that, as $\ell$'
increases, the error for most selectivities first decreases but soon start to rise. This can
be explained by the fact that more restricted privacy (larger $\ell$')
requirement may  compromise the utility. 
For the special case where the query involves only
the sensitive attribute, the relative errors of both small and large counts
 are shown in Figure \ref{fig3}.
The results agree with our analysis in Section \ref{sec:sec6}. The relative error is as well
shown against the selectivity in figure \ref{fig18} .

\subsection{Error for small counts}
We plot the error of queries with small counts separately in Figure \ref{fig2}, where the
counts are smaller than $10$. As one can observe, the error is sufficiently high to
ensure privacy, consistent with our requirement that answer for small count should
be inaccurate enough to prevent privacy leakage. The relative error also displays
a positive linear correlation with $\ell'$. In other words, as $\ell'$ becomes bigger
(higher privacy), privacy for small counts is also ensured at a higher level.


\begin{figure}[tbp]
\begin{center}
\includegraphics[width = 0.9\linewidth,height=1.8in]{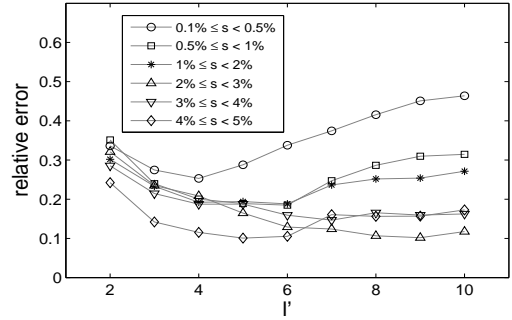}
\end{center}
\vspace{-6mm}
\caption{Relative error}
\label{fig11}
\end{figure}

\begin{figure}[tbp]
\begin{center}
\includegraphics[width = 0.9\linewidth,height=1.7in]{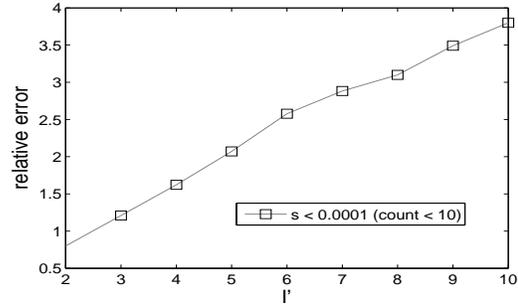}
\end{center}
\vspace{-6mm}
\caption{Relative error for small counts}
\label{fig2}
\end{figure}

\begin{figure}[tbp]
\begin{center}
\includegraphics[width = 0.9\linewidth,height=1.8in]{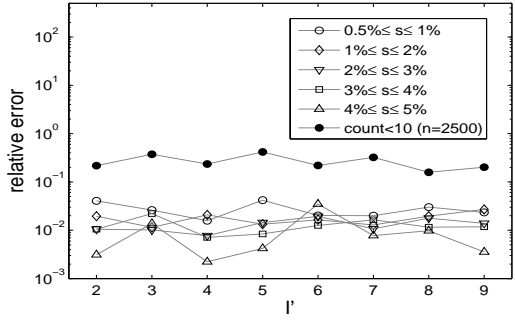}
\end{center}
\vspace{-6mm}
\caption{Relative error versus $\ell'$ for $SA$ querying}
\label{fig3}
\end{figure}

\begin{figure}[tbp]
\begin{center}
\includegraphics[width = 0.9\linewidth,height=1.7in]{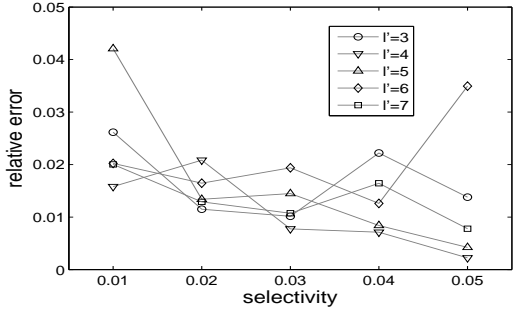}
\end{center}
\vspace{-6mm}
\caption{Relative error versus selectivity for $SA$ querying}
\label{fig18}
\end{figure}

\subsection{Comparison with other models}

To our knowledge there is no known mechanism for $\ell'$-diverted privacy.
We would like to compare the utilities of our method with
other models although it is not a
fair comparison since our method provides guarantees not
supported by the other models.
We have chosen to compare with Anatomy because we have used a similar partitioning
mechanism, and Anatomy is an improvement over previous $\ell$-diversity methods since
it does not distort the non-sensitive values.
We compare with the distortion based differential privacy method since it
has been the most vastly used technique in differential privacy.
Finally we shall compare with the global randomization mechanism $A$
described in Section \ref{init}.
We shall see that our method compares favorably with the other methods
in terms of utility while addressing the dilemma of utility
versus privacy.

To compare with the Anatomy method, we set both $\ell'$ in $\ell'$-diverted and $\ell$ in Anatomy
to the same value, $\ell' = \ell = 5$. The answers for Anatomy are estimated using the method in \cite{XT06b}.
We then choose different $s$ and $N$ (sizes of dataset) to evaluate
their performance. The average relative errors
for Anatomy and mechanism $A'$ are shown in
Figures \ref{fig4} and \ref{fig5}, respectively.
The overall error of our method appears smaller than that
of Anatomy for most choices of $N$ and $s$. The error is bounded by 30\% for 
mechanism $A'$
and can be over 40\% for Anatomy. We can also get an idea of the influence of
different cardinalities of dataset on the error. In fact, the error does not
show an obvious correlation with $N$.

Typical differential privacy secures privacy by adding noises to the answers.
Given a set of queries $q_1, ..., q_m$,
$\epsilon$-differential privacy can be achieved by a randomization
function with a noise distribution of $Lap(\sum_{i=1}^m \Delta f_i / \epsilon)$
\cite{D08}.
Since $m$ is the maximum number of queries that
can be submitted to $D'$, we first set 
$m$ to be 100, and we choose the $\epsilon$ parameter in the Laplacian noise to be 0.01 and 0.05, which are normal choices found in the literature.
The 100k dataset is used, and the average relative error is shown for $s$ between 1\% and 5\% in Figure \ref{fig16}.
The error from differential privacy, no matter which
$\epsilon$ is chosen, will become unacceptably large for smaller $s$.
On the other hand, the impact of $s$ is limited in the case of our method,
the result of which is labeled ``$\ell'$-diverted'' in the graph.
To see how $m$, the number of queries raised, affects the utility, we plot
the relative error against $m$ valued from 10 to 100 in Figure \ref{fig17}.
Obviously the relative error from our method
does not depend on $m$, while that from differential privacy grows linearly
with $m$, and become very large for large $m$.

The results for the global randomization Mechanism $A$ is shown in Figure
\ref{fig16}. We set the value of $p$ to $1/\ell'$ so that the probability
to retain the original sensitive value in each tuple is the same in
both methods.
It can be seen that our method has much better utility for all the
selectivities in our experiment.

\begin{figure}[tbp]
\begin{center}
\includegraphics[width = 0.9\linewidth,height=1.8in]{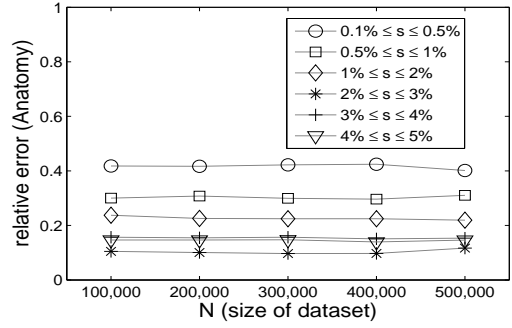}
\end{center}
\vspace{-6mm}
\caption{Utility for Anatomy}
\label{fig4}
\end{figure}

\begin{figure}[tbp]
\begin{center}
\includegraphics[width = 0.9\linewidth,height=1.8in]{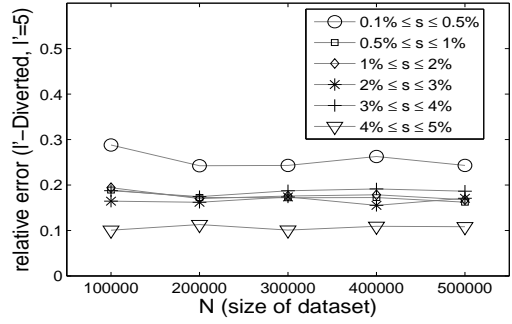}
\end{center}
\vspace{-6mm}
\caption{Mechanism $A'$ for $\ell'$-diverted privacy}
\label{fig5}
\end{figure}


\begin{figure}[tbp]
\begin{center}
\includegraphics[width = 0.9\linewidth]{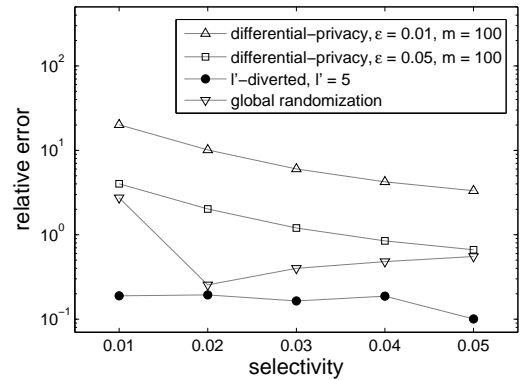}
\end{center}
\vspace{-6mm}
\caption{Comparison of our method ($\ell'$-diverted)
with differential privacy and
global randomization by mechanism $A$}
\label{fig16}
\end{figure}



\begin{figure}[tbp]
\begin{center}
\includegraphics[width = 0.9\linewidth,height=1.8in]{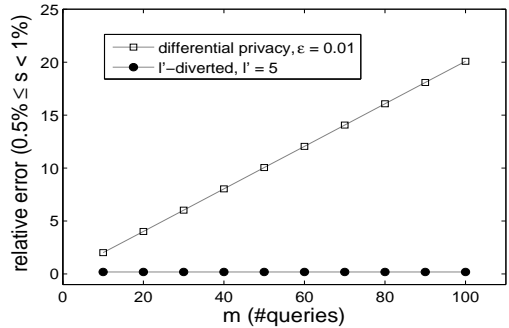}
\end{center}
\vspace{-6mm}
\caption{Multiple queries in differential privacy}
\label{fig17}
\end{figure}


\begin{figure}[tbp]
\begin{center}
\includegraphics[width = 0.9\linewidth,height=1.8in]{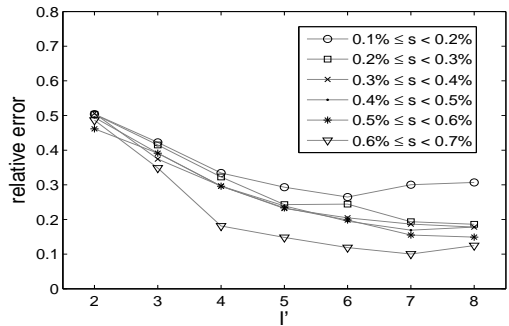}
\end{center}
\vspace{-6mm}
\caption{Relative error for 2 sensitive attributes}
\label{fig172}
\end{figure}
\subsection{Multiple sensitive values}

We also consider the utility in scenarios where a query involves more than one sensitive
value. To this end, we choose \emph{Age} and \emph{Occupation} as the sensitive
attributes. The two sensitive attributes are randomized independently and then combined
for data publication.
To allow queries of large selectivities, we first generalize the domain of \emph{Age}
into ten intervals; without this step, most of the resulting counts are too small and the range
of selectivities is limited. The relative error for multiple-dimension aggregates involving two
sensitive attributes is shown in Figure \ref{fig172}, where $\ell'$ ranges from 2 to 8. Although given
the diminished selectivities ($0.1\%$ to $0.7\%$ for this case), the overall
accuracy can match that in single-sensitive-attribute scenario.

\subsection{Computational overhead}

The computational overhead mainly comes from the partitioning process. We have adopted
the partitioning method of Anatomy. This algorithm can be
implemented with a time complexity of $O(N(1+\frac{V}{\ell'}))$, where $N$
is the cardinality of the table, and $V$ is the number of distinct values of the
sensitive attribute. We show the running time for the case of single sensitive attribute
on the largest 500K dataset, varying  $\ell'$ from 2 to 10.
For all chosen $\ell'$ values, our algorithm can finish within 10 seconds for a 500K dataset, which
is practical to be deployed in real applications.

We also consider the querying efficiency at the user side. To estimate the answer, a user will
compute each component of the vector $y$, and do matrix multiplications to iteratively
converge at the answer $x$. When each component of $y$ changes by no more than $1\%$,
we terminate the iteration and measure the querying time and number of iterations.
In our experiments, SQLITE3\footnote{See http://docs.python.org/library/sqlite3.html} serves for querying $y$, and we consider the case with two sensitive
attributes which involves the most number of components in $y$, implying the largest computational
cost. The result shows that the Bayesian iterative process takes negligible time, while the major
cost comes from the querying step. In particular, it takes less than 1 ms in average, and 10 ms
in the worst case, for the iterative process to converge. The median and average of the number
of iterations is 16 and 325, respectively. In total, the average measured time for a query is
1612 ms, which poses little computational burden on users.

\section{Related Work}
\label{related}

Differential privacy has been a break-through in the study of
privacy preserving information releases.
$\epsilon$-differential privacy has been introduced for query answering
and the common technique is based on
distortion to the query answer by a random noise that is i.i.d. from a
Laplace distribution and calibrated to the sensitivity of the querying
\cite{DMNS06,D06}.
Laplace noise has been used in many related works on differential
privacy including
recent works on reducing relative error
\cite{X11} and the publication of data cubes in \cite{DWHL11}.
Since the data release can be for different purposes,
in some tasks, the addition of noise makes no sense. For example, a utilization function might map databases to strings, strategies, or trees.
The problem of optimizing the output of such a function while preserving
$\epsilon$-differential privacy is addressed in \cite{MT07}.
%
For database publication, \cite{BLR08} shows that given a large enough
dataset, a synthetic database can be generated that is approximately correct for all
concepts in a given concept class; the minimal data size depends on
the quality of the approximation, the log of the size of the universe, the privacy parameter $\epsilon$ and the Vapnick-Chervonenkis dimension of the concept class.
%
%
Further results can be found in
\cite{DNRRV09}.
In most previous works, the definition of error is an absolute error
\cite{DN03,DMT07,BLR08,DNRRV09}.
The algorithm iReduct in \cite{X11}
considers relative errors and injects noise to
query results according to the values of the results.
A recent work \cite{KM11} points out that differential privacy
may not guarantee privacy when deterministic statistics have been
previously published. In contrast we consider a more
basic possible privacy leak which is due to the fact that
differential privacy does not aim to protect
information that can be derived from the published data,
deeming such a task impossible.
All previous works on differential privacy
consider $\epsilon$-differential
privacy for non-zero $\epsilon$ values.
None of the works in the above considers the guarantee of
protection of small sums, which is a major objective in our mechanism.

In the literature of statistical databases,
the protection of small counts has been well-studied in
the topic of security in statistical databases
\cite{AdamSecurityControl89}.
A concept similar to ours is found in
\cite{TYW84} where the aim is to ensure that
the error in queries involving a large number of tuples will
be significantly less than the perturbation of individual tuples.
It has been pointed out in previous works
\cite{Fellegi72,HM76} that the security of a database is endangered
by allowing answers to counting queries that involve small counts, i.e. the number of tuples involved in the query is small.
In \cite{Denning80}, random sampling has been
used to ensure large errors for small query set sizes.
However, these previous works are about the secure disclosure of statistics
from a dataset and
do not deal with the problem of sanitization of a dataset for
publication, and they have not considered the guarantee of differential privacy.
Discriminative privacy protection has been considered in some previous work
in privacy preserving data publication such as
\cite{XT06a,YCY10}, however, such works are based on personalized privacy requirements.
There have been studies that the utility of published dataset can lead to privacy breach
\cite{Kifer09,WFWXY11}, however, they focus on
partition-based methods for $\ell$-diversity and they have
pointed out the problems while no solution is proposed.

Randomization technique has been used in previous works in privacy preservation.
The usefulness of such a technique is shown in
\cite{agrawal-preserving-data-mining}, where the published data is used to build a decision
tree which achieves classification accuracy comparable to the accuracy of
classifiers built with the original data. An effective reconstruction method for data
perturbation is introduced in
\cite{AgrawalPODS01}. In \cite{agrawal05}, random perturbation is adopted for
privacy preserving computation for multidimensional
aggregates in data horizontally partitioned
at multiple clients.
Randomization of transaction datasets for the mining of
association rules has been considered in \cite{ESAG02}.

\section{Conclusion}

We have introduced a new mechanism in the problem of
privacy preserving data publication with the following
properties. Firstly, it satisfies
$\ell'$-diverted zero-differential privacy, which makes sure
that the resulting data analysis will have no
 difference whether an individual keeps its true sensitive
 value or swap the
true value with other individuals.
Secondly, the randomization process makes use of the law
of large numbers in ensuring that large counts, which
are not as sensitive, can be estimated with high accuracies
while the small counts will be hidden by relatively large errors.
Our method is parameter free except for the value of $\ell'$,
however, the choice of $\ell'$ has little effect on the privacy and
as shown in our experiments, setting $\ell'$ to 5 or above
will do well in terms of the utilities.
Furthermore, the sensitive value of a tuple in the
published data can be any value in the attribute domain, so
the mechanism is resilient to auxiliary knowledge
which eliminates possible values.
Our empirical studies on a real dataset show superior
utility performance compared to other state-of-the-art
methods which do not have the above guarantees.
For future work, we may consider how to handle
skewed sensitive microdata
\cite{XWF+10}.
Another direction for future work is to consider mechanisms such
as small domain randomization for further boosting the utilities
for large counts \cite{CW10}.
The consideration of sequential data releases is another open
problem.

As a final remark, all existing privacy models inherently release
information that can be derived from the published datasets, and
the same is true with our approach. It is important to make known
to the users what kind of information they should expect to be
released or derivable. In our case, it will be relatively
accurate answer to queries with large sums.

\bibliographystyle{abbrv}
\bibliography{reference}


\end{sloppy}
\end{document}